\documentclass[11pt,a4paper]{article}
\usepackage{jinstpub}
\pdfoutput=1

\usepackage{hyperref}

\usepackage[english,multiuser,final]{fixme}
\fxusetheme{colorsig}
\fxloadtargetlayouts{colorcb}
\fxusetargetlayout{colorcb}
\fxuselayouts{margin,pdfnote}
\FXRegisterAuthor{ed}{editor}{editor}
\FXRegisterAuthor{dw}{danw}{danw}

\title{An Electro - Optical Test System for Optimising Operating Conditions of CCD sensors for LSST}
\author[a,1]{D. P. Weatherill \note{corresponding author}}
\author[a]{K. Arndt}
\author[a]{R. Plackett}
\author[a]{and I .P. J. Shipsey}

\affiliation[a]{University of Oxford, Keble Road, Oxford OX1 3RH, UK}

\emailAdd{daniel.weatherill@physics.ox.ac.uk}


\abstract{We describe the commissioning of a system which has been built to investigate optimal operation of CCDs for the LSST telescope. The test system is designed for low vibration, high stability operation and is capable of illuminating a detector in flat-field, projected spot, projected pattern and Fe-55 configurations. We compare and describe some considerations when choosing a gain calibration method for CCDs which exhibit the brighter-fatter effect. An optimisation study on a prototype device of gain and full well with varying back substrate bias and gate clock levels is presented.}

\begin{document}
	\listoffixmes
	\maketitle
	\flushbottom
	
	\section{Introduction}
	In this contribution we present the first results from the commissioning and early runs of an electro-optical test system which will be used to investigate optimisation of the operating parameters of CCDs for use in LSST, and also to investigate charge collection and transfer effects which may be of interest with regard to the LSST science goals (see \cite{Abell2009} for a detailed description of LSST science requirements). The specifications and standard test procedures for LSST sensors are not described here (see \cite{doi:10.1117/12.2056733} for fuller detail). Of particular interest to the subject of calibration, is that the LSST devices are highly segmented compared to many other CCDs (16 outputs), that the readout is fairly quick (550kHz pixel rate); and that the thick, back biased nature of the sensor leads to quite high non-linearities in charge collection (known widely as the "brighter-fatter" effect \cite{1748-0221-9-03-C03048}). In the following sections we describe the design and construction of the test system, and present initial results from studying mean variance curves for various CCD bias conditions.
	
	\section{Experimental Setup \label{expt_setup}}
	The CCD test stand has been designed to be complementary to other facilities available to the LSST project (for a description of the test facility used in acceptance testing and benchmarking of sensors, see \cite{doi:10.1117/12.2231925}).
	In particular, attention has been paid to low vibration operation and time stability. 
	
\begin{figure}
		\includegraphics[scale=0.3]{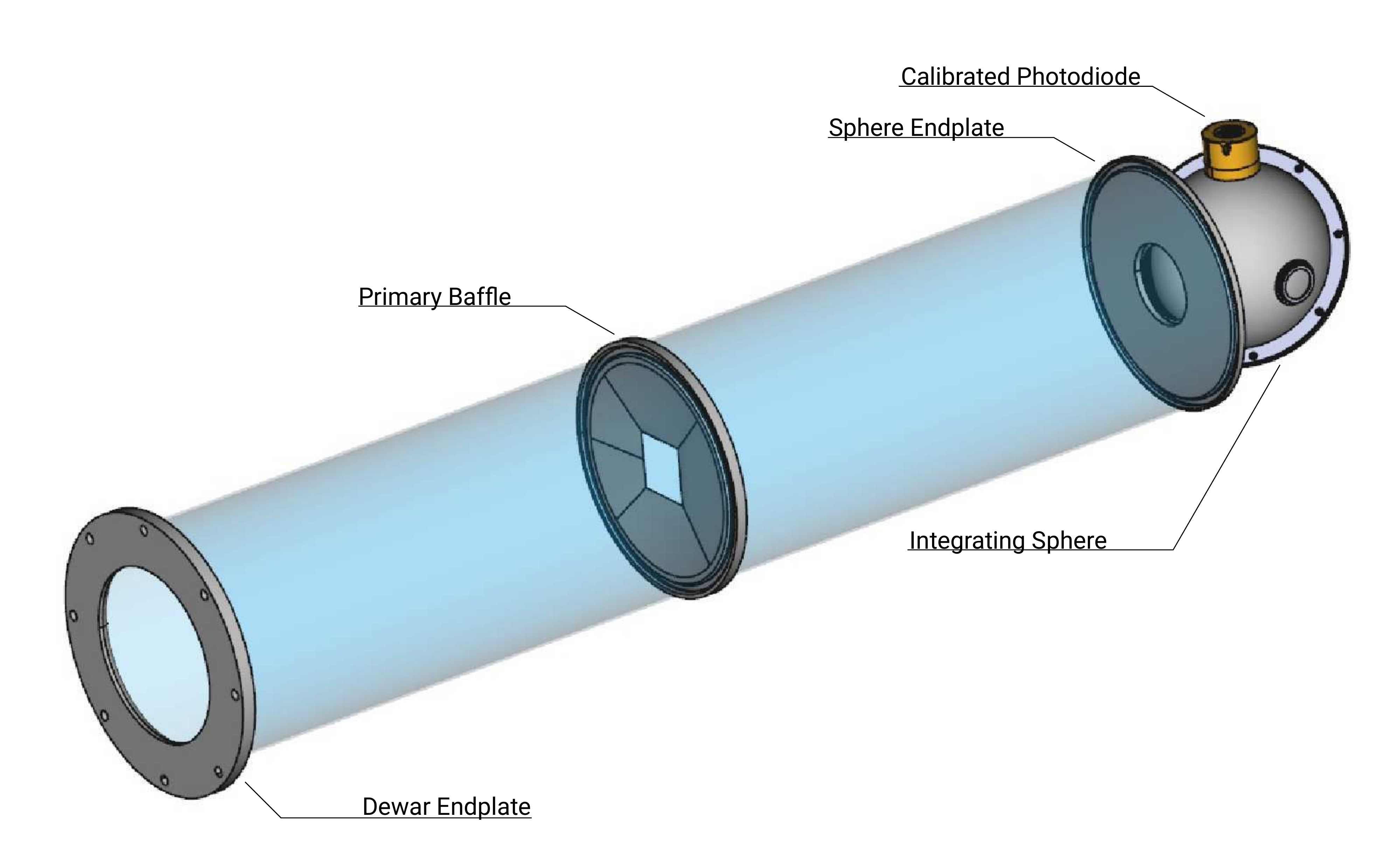}{\centering}
		\caption{CAD drawing of baffle configuration\label{baffle_config}}
	\end{figure}

	A 250W Quartz Tungsten Halogen (QTH) lamp illuminates a monochromator via an order sorting filter. The output is coupled via a focussing lens assembly and liquid light guide to a 6" integrating sphere. A calibrated photodiode monitors one port of the integrating sphere at all times, and a visible spectrometer can be attached to another port in order to calibrate the spectral content of the system. A custom baffle tube of 1m length is installed at the main output port of the integrating sphere. The tube contains a single rectangular knife-edge baffle, and the geometry of both tube and baffle are chosen such that all stray diffuse rays arising from up to one reflection are removed (see \autoref{baffle_config}). Inside, the tube is coated in black flocked paper to eliminate specular reflection and reduce diffuse reflection as far as possible. The end and centre plates of the baffle tube were manufactured in matte nylon using 3D printing.

	When spot or pattern illumination is required, the baffle tube can be removed, and a projector system can be slid in front of the cryostat along precision dovetail optical rails. The projector system consists of a small (2") integrating sphere, an iris aperture, a light tight target wheel, and a matched achromat doublet lens pair. The whole assembly is mounted on an XYZ positioning stage, with $1 \mathrm{\mu m} $ resolution in the CCD array directions, and $100 nm$ resolution in the focussing direction. The travel of the stages is such that the entire area of the CCD ($\approx 45 \mathrm{mm}$ square ) can be covered. As with the flat field system, illumination is introduced via liquid light guide. When used in this mode, a large dark enclosure is placed over the entire system, and a small conical baffle is attached to the cryostat window.

\begin{figure}
 		\includegraphics[scale=0.4]{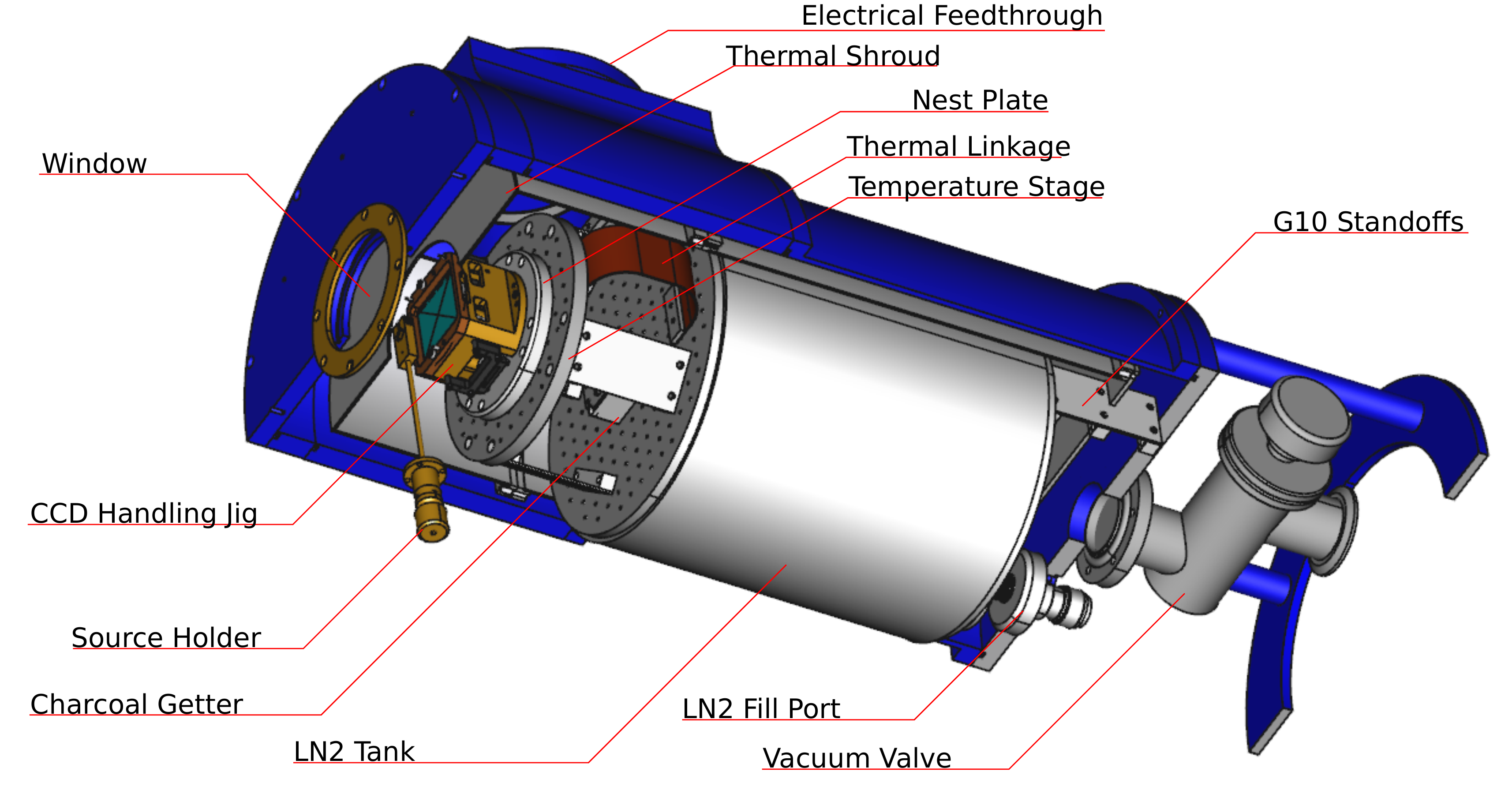}
 		\caption{Partial cutaway view of the testing cryostat \label{cryo_config}}
 	\end{figure}

	The CCD under test itself is housed in an aluminium cryostat (manufactured by Universal Cryogenics), which is evacuated to high vacuum using a turbopump, and can then be subsequently isolated via a valve and the pump deactivated to reduce vibration during measurement. A 6L tank is situated behind the temperature stage within the cryostat, which can be filled externally with liquid nitrogen. Operating temperature is maintained with a PID controlled 25W cartridge heater. The filling procedure is automated using a cryogenic solenoid valve attached to a self-pressurising storage dewar, and occurs roughly daily.
	 The cooling system is thus virtually vibration free outside of these brief filling cycles, especially compared to an active compressor based continuous cycle system. Six 50-way D-SUB connectors provide electrical feed-through into the cryostat, and a UV-grade fused silica window allows illumination. A manually actuated linear rod enables a small Fe-55 radioactive source to be positioned in front of the CCD. A schematic of the cryostat configuration is shown in \autoref{cryo_config}.
	 
	CCD readout and clocking is controlled by an STA Archon\cite{doi:10.1117/12.2058402}.
	This provides 16-channels of digital CDS readout running at 100MHz, clock drivers for the image, register and reset clocks, and DC voltages for the output drain ($V_\mathrm{OD}$), output gate ($V_\mathrm{OG}$), reset drain ($V_{RD}$), guard drain ($V_{GD}$) and back substrate ($V_{BB}$) bias lines. Preamplifiers and differential converters on custom boards are attached directly to the cryostat feedthrough (these are of the same design as described in \cite{doi:10.1117/12.2231925}), and signals, clock lines and biases are passed to a custom archon adaptor board through dual link DVI-I cables, which are used due to the high number of differential pairs available at low cost.
	
\begin{figure}[h]
		\includegraphics[width=\textwidth]{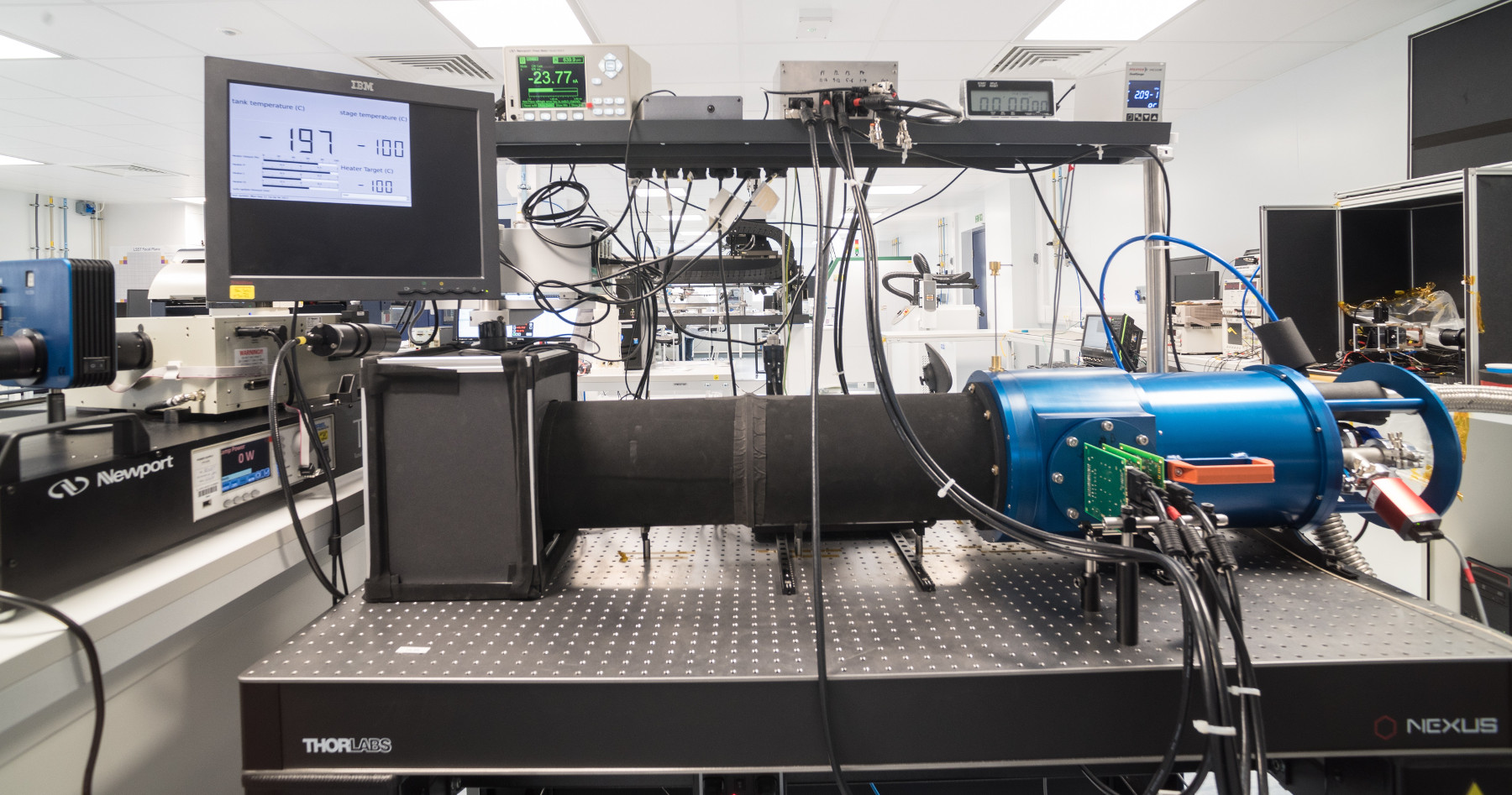}{\centering}
		\caption{Photograph of optical test system in flat field configuration \label{test_system_photo}}
		\ederror{capitalise first letter and axes labels}
	\end{figure}

	The entire system apart from the light source and monochromator is mounted on an active pneumatically controlled optical table of $ 900 \mathrm{mm} \times 1200 \mathrm{mm} $. As far as possible, vibration causing equipment (e.g. containing fans) is kept off the table. A photograph of the whole system is shown in \autoref{test_system_photo}.

	\section{Gain and Full Well Measurements}
	Accurate and robust measurement of the camera gain is critical, not least because many other performance measurements depend directly on knowledge of the gain. In most CCDs, measuring gain, full well, read noise and other quantities is accomplished using the photon transfer curve (based on shot noise statistics of incident photons), or else using an Fe-55 radioactive source (illuminating with X-ray photons of a known energy) \cite{janesick2001scientific}. In a thick, back illuminated, deep depletion device such as those used for LSST however, both methods present some complications.
	
	Due to the brighter-fatter effect, flat field illumination of the CCD results in correlations between nearby pixel values which are not caused by traditional crosstalk mechanisms in the readout chain \cite{1748-0221-9-03-C03048}. In terms of photon transfer analysis, this exhibits as a non-linearity in the expected linear mean-variance curve \cite{doi:10.1117/12.671457}. One approach to obtain the gain information in these circumstances is to fit a quadratic, rather than linear, curve to the mean variance data \cite{6681961,doi:10.1117/12.2024263}. Another is to recover the linearity by constructing a mean-covariance curve where nearby pixel to pixel co-variances are added to the variance \cite{guyonnet2015evidence}. Both methods are briefly examined below. Measuring the gain using Fe-55 illumination is complicated by the thickness of the device (100 $\mu$m), and the small pixel size (10 $\mu$m): since the vast majority of X-ray photons convert very near the back surface of the device and must drift to the collecting gates, the resulting electron clouds are found to be mostly split over multiple pixels.

	We now turn to discussing the implementation of each method. The mean variance data in each case is obtained from difference pairs of flat field illuminated frames, after serial overscan correction (which is performed using a single median average of the serial overscan pixels). Each difference image is then masked for cosmic rays using a local median filter method (connected groups of more than 2 pixels which are greater than 5 times the image standard deviation $\sigma$ above the median level of the nearby 10 square pixel region are masked). Any difference images with a median more than 50DN away from zero are excluded (arising for example due to light source instability), though these are rare cases. Implementations of the fitting, minimisation and interpolation procedures used are those from scipy \cite{4160250}.

		\subsection{Fe-55 calibration}
\begin{figure}
			\includegraphics{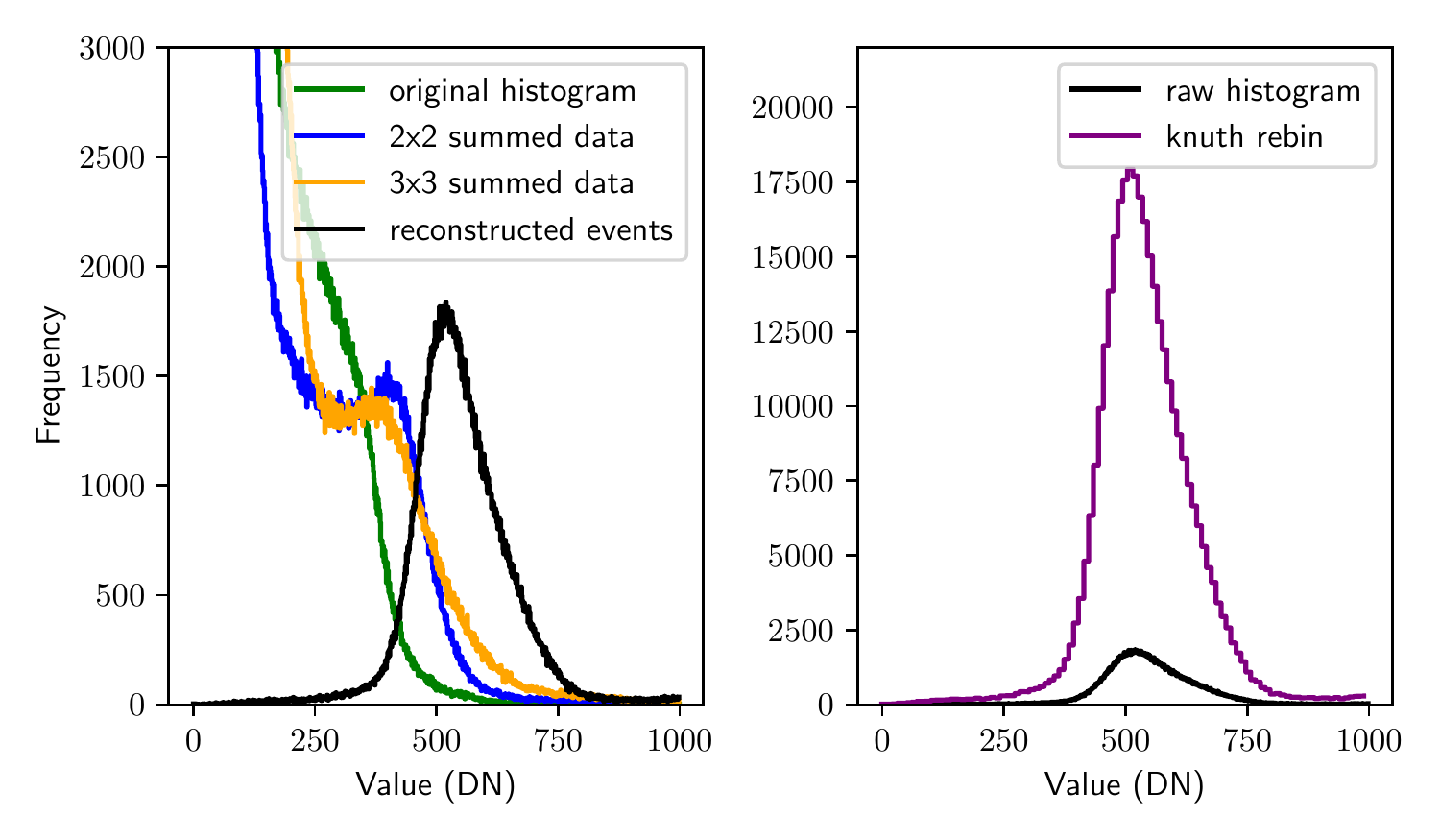}
			\caption{example Fe-55 histograms, constructed from 100 images, with $V_\mathrm{BB}=-60\mathrm{V}$. We compare raw images, summed in 2x2 and 3x3 regions, and simple event processed data (left panel). We demonstrate the use of optimal re-binning via Knuth's rule to improve histogram quality (right panel)\label{fe55_histograms}}
			\ederror{capitalise first caption letter and axes labels}
		\end{figure}

		The standard method to measure gain using Fe-55 illumination is to find the peak due to Mn-K$\alpha$1 absorption line (at 5898 eV) in the image histogram, which deposits a known number of electrons ($~1615$) into the pixel. As previously noted, however, the device thickness and small pixel size results in very few single pixel collections of x-ray charge clouds. To recover the energy peak, the image pixels can be summed over small areas, but in our case even this does not accurately recover the peak. Ideally, this process would be performed using on-chip binning (to eliminate the extra read noise cost in summing the pixels after readout), though this was not performed in our investigation. A simple event recognition algorithm (based on finding values more than 5$\sigma$ above the background image level and summing the nearby 3x3 region) produces better results (see \autoref{fe55_histograms}). A similar approach has been used by many previous authors, e.g. \cite{doi:10.1117/12.876627}. The histograms shown are "raw", in the sense that they use 1 bin per DN value in the original image. The use of Knuth's procedure for optimal re-binning of histograms \cite{2006physics...5197K} can then be used with great effect to improve the visibility of features (see right panel of \autoref{fe55_histograms}). We are unable to see the \ednote*{use greek letter}{Mn-K$\beta$} peak distinctly than that of Mn-K$\alpha$, which is likely because the level of \ederror*{please state level of read noise in the system}{read noise in our system is presently not low enough.}
	
	\subsection{Finding Full Well \label{find_full_well}}
		\begin{figure}
			\begin{center}
				\dwerror{units on derivative axis}
				\includegraphics{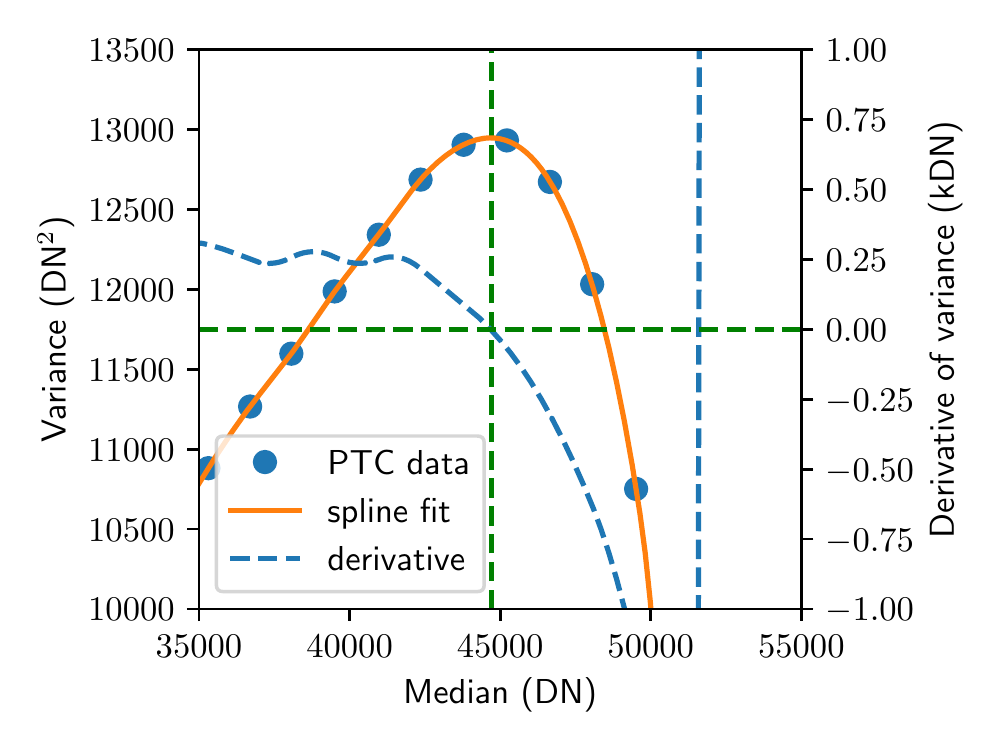}
				\caption{finding full well via a spline fit \label{find_full_well_graph}}
				\ederror{capitalise first caption letter and axes labels}
			\end{center}

		\end{figure}
		In order to successfully use either of the possible non-linear fitting methods, it is required that a good estimate of the full well is first obtained, because above this point, the mean-variance curve changes behaviour. Simultaneously, a good measurement of full well is often \ednote*{confusing, please re-word}{desired knowledge in any case}. To find full well from the mean variance data, first we find the point at which the maximum variance $\sigma^2_\mathrm{max}$ was obtained. Since it is unlikely that a data point we have taken in the experiment falls exactly at full well, we then fit a cubic spline representation to the data points which lie either side of the maximum, and which have a variance value $\sigma^2 \geq {\sigma^2_\mathrm{MAX} \over 2}$. We then use Brent's method \cite[Chapter 4]{brent2013algorithms} to find the turning point (via finding the root of the spline's derivative), and call this value the full well capacity ($s_\mathrm{FW}$). This procedure is illustrated in \autoref{find_full_well_graph}. Note that in general the full well capacity defined in this way is significantly different than the saturation level of the output amplifier (as shown in \autoref{ptc_schematic}), though both are important quantities which can be optimised using bias conditions and clocking.
	
	\subsection{Quadratic Fitting \label{quadratic_fitting}}
		\begin{figure}
			\includegraphics{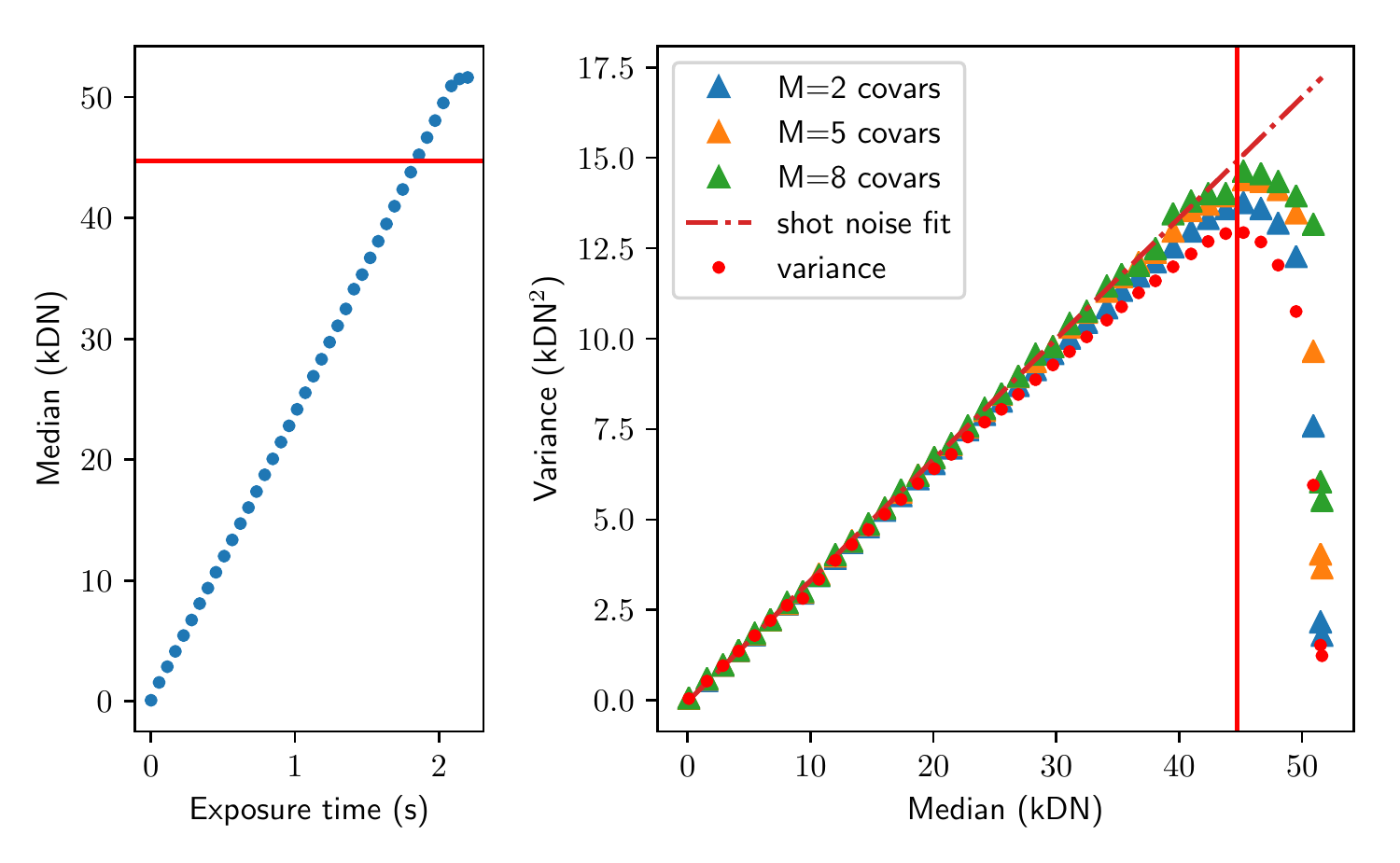}
			\caption{example of a mean variance curve for one CCD output channel. Red points show the mean-variance data, triangular marks show the reconstructed mean-covariance data (see \autoref{mean_cov}) for various covariance distances $M$. The equivalent shot noise line from a  quadratic fit is shown by the dashed red line. Note that the full well capacity identified (red vertical line in right hand panel, horizontal line left hand panel) is not the same as the level at which the output signal saturates (as seen in the left hand panel) \label{ptc_schematic}
			}
						\ederror{The quadratic fit does not seem to fit well the variance (red circles) - please explain why. Also quantify difference between FWC calculated from peak variance and signal saturation. Capitalise caption first letter and axes labels}

		\end{figure}
		Following \cite{guyonnet2015evidence}, we fit the following function (via a least squares polynomial fit) to the mean variance data:
		\begin{equation}
			\sigma^2\left(\bar{s}\right) = -{1\over 2}\alpha \bar{s}^2 + {\bar{s} \over {2K}} + {\sigma^2_0\over 2}
		\end{equation}
		
		where $\bar{s}$ is the median signal level (in digital number units), $\sigma^2_0$ is the offset variance, $\alpha$ is the non-linearity and $K$ is the camera gain. The factor of $1\over 2$ is necessary because in combining two separate images and taking the difference, we have doubled the variance due to shot noise \cite{janesick2001scientific}. Only those data where $ S \leq \gamma S_\mathrm{FW}$ are used, where $\gamma$ is a somewhat arbitrary choice, though clearly it must be less than unity (see \autoref{find_full_well}). In practice, we find a value of $\gamma=0.9$ to be robust over many mean-variance curves. It is,however, found that the choice of $\gamma$ can influence the resulting gain more than the reported fit error, which is a problem if agreement on the exact gain value is important (see \autoref{gain_comparison}). A sample fitted quadratic and corresponding derived linear gain curve (i.e. with slope $1 \over K$) is shown in \autoref{ptc_schematic}). 
		
	\subsection{Mean - Covariance Curve \label{mean_cov}}
		The brighter - fatter effect causes electrons correlations between nearby pixel values, and thus reduces the image variance. This "missing" variance is effectively moved to the covariance between pixels. Defining the covariance between pixels in a difference image $C_{n,m}$ as:
		\begin{equation}
			C_{n,m} = { \sum_{i,j} s_{i,j}s_{i+n,j+m}  \over \sum_{i,j} s^2  }
		\end{equation}
		where the sums run over all the image pixels, and $s_{i,j}$ represents the value of the pixel at position $\left(i,j\right)$. The covariance sum $D(M)$ is given by:
		\begin{equation}
			D(M) = \sum_{n=1,m=1}^M C_{n,m}
		\end{equation}
		The mean-covariance curve is then constructed:
		\begin{equation}
			{\sigma^2}^\prime\left(\bar{s}\right) = {\sigma^2\left(\bar{s}\right) \over {2K}} + {D(M)\sigma^2\left(\bar{s}\right)\over {2K}} + {\sigma^2_0\over 2} 
		\end{equation}
		As $M$ is increased, linearity of the gain curve is gradually recovered. Mean-covariance curves for several values of $M$ are shown in \autoref{ptc_schematic}. A linear function can then be fitted to the mean-covariance data. In common with the quadratic fitting case, $\gamma$ must be chosen suitably, though we find a wider range of values over which the resulting gain measurement is stable (see \autoref{gain_comparison}). In addition, a value for $M$ must be chosen. In principle, the higher the value for $M$ the better, since brighter fatter correlations can extend over many pixels. However, factors such as read noise and true cross-talk introduced by downstream readout electronics will eventually become dominant over brighter-fatter correlations at high $M$.

	\subsection{Comparison of Gain Methods \label{gain_comparison}}
 		\begin{figure}[pbth]
 			\includegraphics{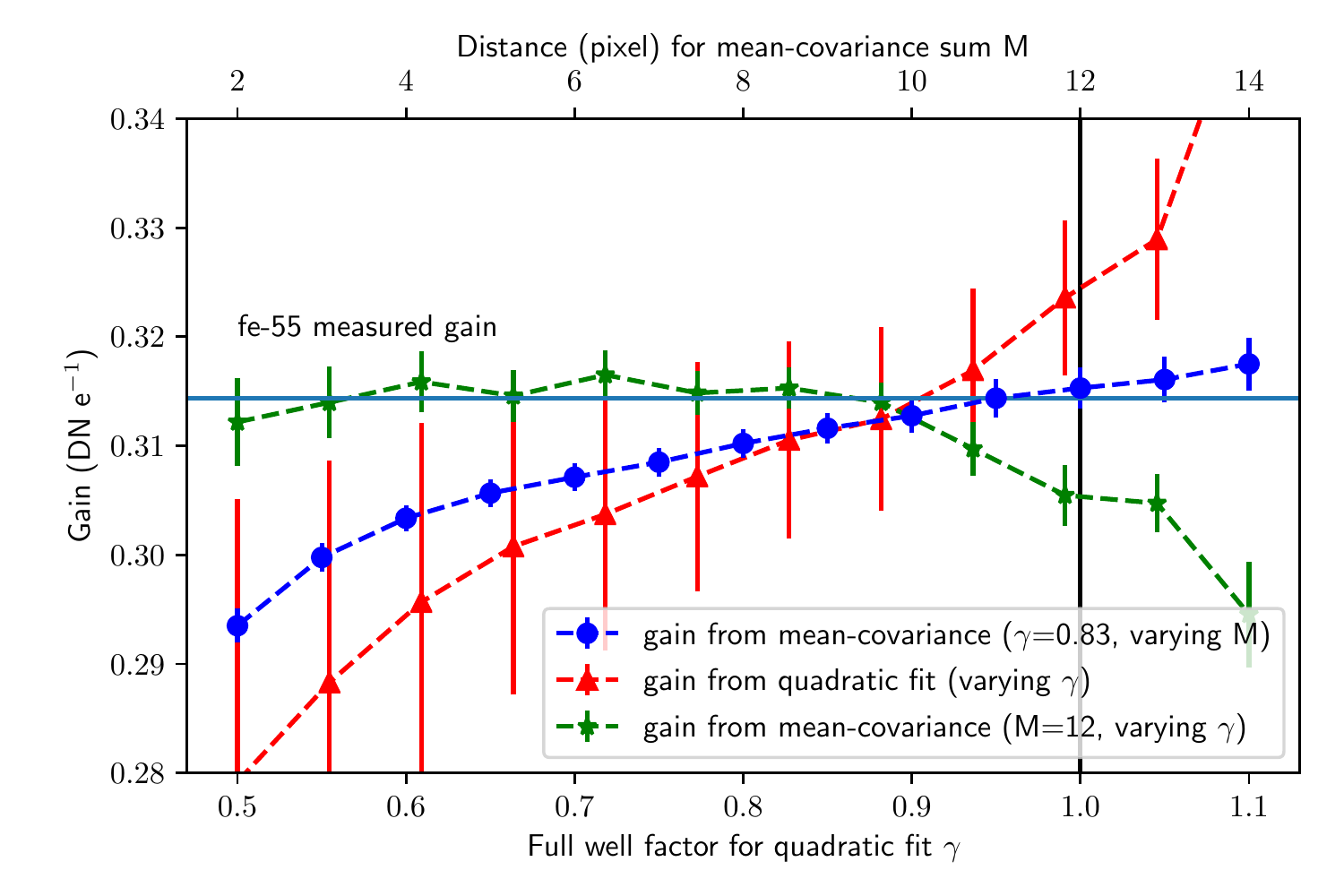}
 			\caption{Measuring the camera gain using different methods from the same mean-variance data. In each case, errors are derived from the fitting covariance. The horizontal line shows gain as measured by Fe-55 for the same channel. \label{ptc_comp_graph}}
 			\ederror{capitalise axes labels}
 		\end{figure}	

		An example of finding the camera gain using the methods described in \autoref{quadratic_fitting} and \autoref{mean_cov} for a particular CCD channel are shown in \autoref{ptc_comp_graph}. The agreement between mean-variance fitting methods and Fe-55 increases with $\gamma$, though as expected, becomes less accurate with $\gamma \geq 1$. Agreement with Fe-55 measurement also increases in the case of the mean-covariance method with increasing $M$, though beyond a certain point the derived values over-estimate with respect to the Fe-55 gain, possibly because any correlated external noise sources with a comparatively low frequency begin to contribute at higher distances.

	\section{Voltage Level Optimization}
\begin{figure}[pbth]
		\includegraphics{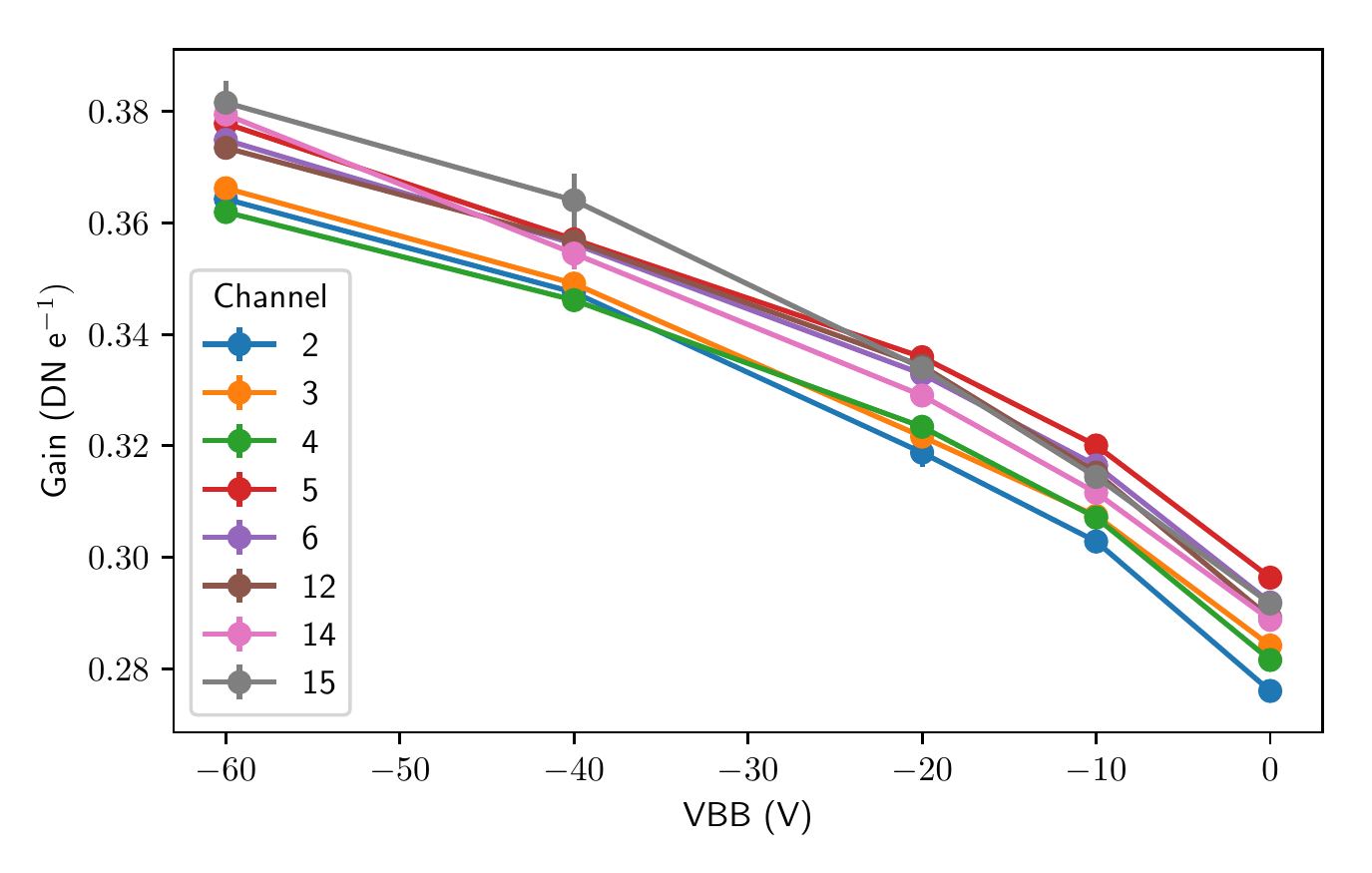}{\centering}
		\caption{Variation of measured output gain as a function of backside substrate bias, $V_\mathrm{BB}$. The measurements displayed here use the mean-covariance method with $\gamma=0.8$ and $M=12$ \label{gain_vs_vbb_graph}}
		\ederror{capitalise axes labels}
	\end{figure}

	The gain value measured as the backside bias of the CCD varies has been observed to change by Robbins et al in previous work \cite{doi:10.1117/12.876627}. The most likely physical reason is the dependency of the depletion capacitance of the pn junction forming the sense node on the bias. We also observe this behaviour (see \autoref{gain_vs_vbb_graph}), though it seems that the strength of this effect is much greater in the LSST CCD than for the (in some ways similar) device examined by the previous authors' work. 
	
\begin{figure}
		\includegraphics{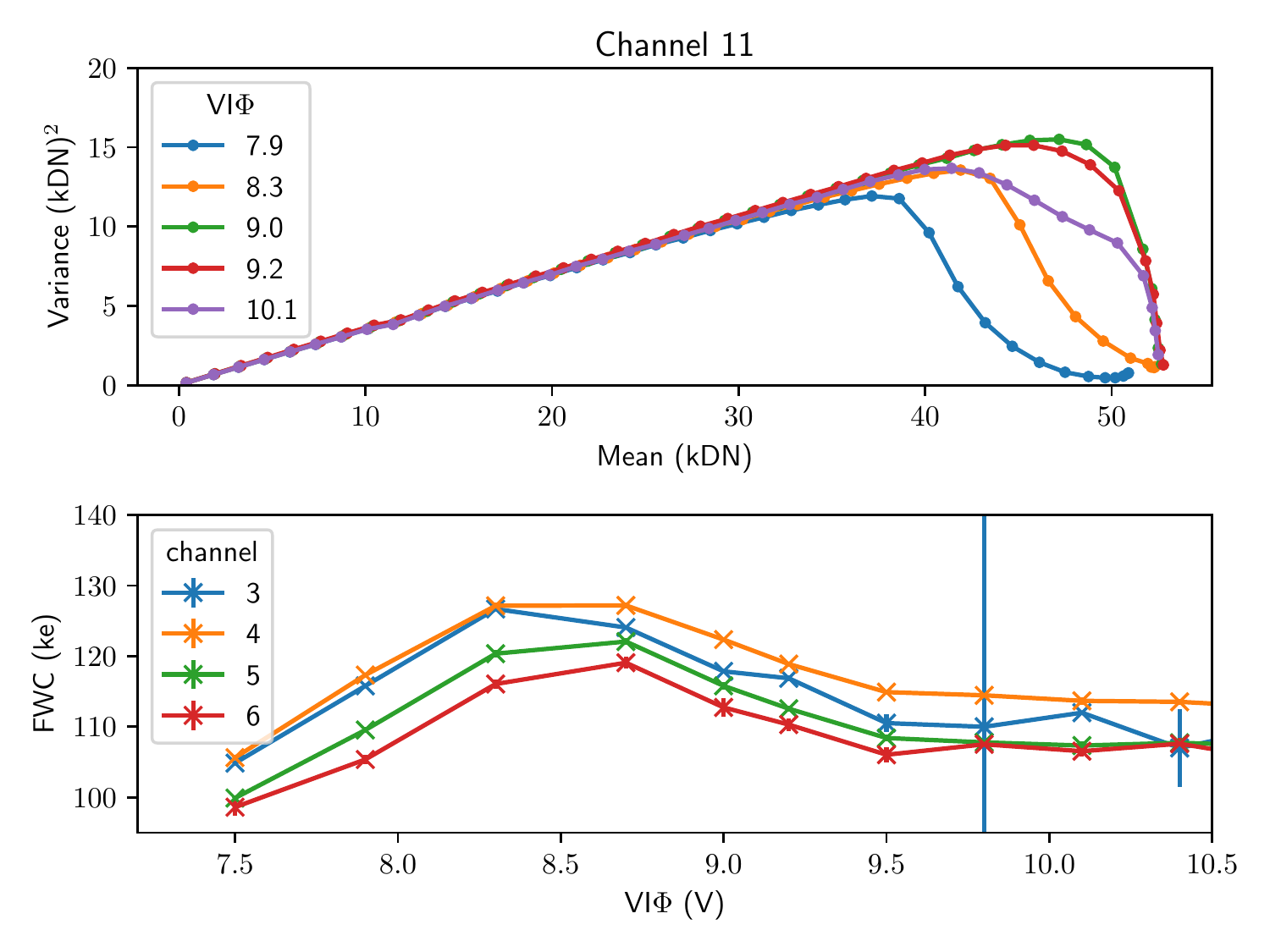}
		\caption{Measurement of full well as integration clock voltage $V_\mathrm{I\Phi}$ is altered. (above panel) the shape of mean-variance curves for various image clock voltages. (below panel) The measured full well capacity as image clock is adjusted. The vertical blue line shows the measured inversion threshold potential \label{iphi_fw_graph}}
		\ederror{capitalize axes labels}
	\end{figure}

	An important consideration in choosing the operating image clock levels of a CCD is to prevent surface channel transport. As a comparison, we first measured the channel inversion threshold using the method described by Janesick \cite{janesick2001scientific}: finding the level of $V_\mathrm{RD}$ at which charge injection into the output node begins. This value was found to be $V_\mathrm{INV}=(9.8 \pm 0.1) \mathrm{V}$. One would expect the optimum value of the image clock in terms of full well capacity to be slightly below this value. The forms of some mean variance curves, and the resulting measured full well capacities for selected CCD channels are shown in \autoref{iphi_fw_graph}. 
	Indeed the optimum full well is found to be approximately 1.25V below  the inversion threshold. Note also that above the optimum image clock operating point, the shape of the mean variance curve above full well is found to be qualitatively different, the variance dropping more gradually before saturation.
	The value of the full well capacity of different channel varies somewhat significantly, though all channels appear to agree well on the position of the optimum image clock value.
	
	\section{Conclusions}
	An experimental system has been described which is designed to assist with optimisation studies of charge collection and operating points of CCD sensors for LSST.
	
	An investigation into different ways to calibrate gain in the presence of brighter-fatter effect has been conducted, and it is found that gain measurements using non-linear techniques on mean variance data agree with those from Fe-55, though only with some selections of the data used in the fit. Our recommendation at this stage would be in general to attempt calibration using the mean-covariance method, which appears to be more broadly stable over a broader range of chosen parameters. 
	
	We measured gain and full well on a prototype LSST sensor, whilst varying the backside bias and image clock voltage levels. To optimise full well and avoid surface transport effects, the image clock must be set to a value lower than the channel inversion threshold. Indeed, to retain charge transfer efficiency performance this level should be chosen rather conservatively.
	
	\acknowledgments
	We wish to acknowledge funding for this work from STFC and Oxford University. Many thanks to colleagues at Brookhaven National Laboratory for their support and useful discussions throughout the work. Chris Damerell at Oxford has also provided much useful input and discussions.

	\ednote{All references to be in JINST format}
	\bibliography{references}
	
\end{document}